\documentclass{jfm}

\usepackage{graphicx}
\usepackage{newtxtext}
\usepackage{newtxmath}
\usepackage{natbib}
\usepackage{hyperref}
\hypersetup{
	colorlinks = true,
	urlcolor   = blue,
	citecolor  = black,
}

\newcommand{\RomanNumeralCaps}[1]

\usepackage{subfigure}
\usepackage{todonotes}

\usepackage{xparse}

\newcommand{\dd}[1]{\mathrm{d}#1}

\newcommand{\pder}[2]{\frac{\partial#1}{\partial#2}} 
\newcommand{\tder}[2]{\frac{\dd#1}{\dd#2}} 
\newcommand{\pderb}[2]{\frac{\partial#1}{\partial#2}}

\newcommand{\delforsimp}[1]{}

\title{Variational determination of minimal absorbing zones in incompressible shear flows}

\author{P\'{e}ter Tam\'{a}s Nagy\aff{1}
	\corresp{\email{pnagy@hds.bme.hu}},
}

\affiliation{\aff{1}Department of Hydrodynamic Systems, Faculty of Mechanical Engineering, Budapest University of Technology and Economics. M\H{u}egyetem rkp. 3., H-1111 Budapest, Hungary
}
\begin{document}
	\maketitle
	
	\begin{abstract}	
The dynamical analysis of shear flows remains challenging, as turbulence generation and evolution are not fully understood. Here, a lesser-explored feature of incompressible shear flows-the “absorbing zone”- is investigated. This region in the infinite dimensional state space is shown to act as an attractor for all trajectories: any solution initialised outside eventually enters and remains inside. Consequently, the absorbing zone must contain all possible attractors, both chaotic and non-chaotic.

Existence is established through the Reynolds-Orr identity, which indicates that nonlinear terms do not directly influence the temporal evolution of kinetic energy. The zone is constructed around a so-called shift flow, and multiple such regions may exist, even when the laminar state is linearly unstable. Gradient based optimisation is employed to identify the absorbing zone of minimal radius. Assuming a chaotic trajectory explores state space randomly, it is hypothesised that the centroid of this minimal zone approximates the turbulent mean flow. This central state is computed for plane Poiseuille and Couette flows and compared with established turbulent mean profiles.

Although the proposed hypothesis is found to be an oversimplification-yielding profiles that qualitatively resemble but do not quantitatively reproduce the turbulent means-the methodology provides novel insights into shear flow dynamics. Furthermore, it offers a promising foundation for refining estimates of global stability thresholds. With continued development, this framework may facilitate the direct computation of turbulent mean states without reliance on empirical turbulence models or time-dependent numerical simulations.		
	\end{abstract}

	\section{Introduction}
	\label{sec:introduction}

	The Navier-Stokes equations, cornerstone of fluid dynamics, represent a quintessential example of a non-linear system capable of exhibiting remarkably complex behaviour \citep{Doering2009, Jimenez2018}, including the transition from laminar to turbulent flow and fully developed turbulence itself. Understanding the intricate phase space structure of such systems is paramount for prediction and control. Over the past decades, a dynamical systems approach has provided profound insights, revealing that the seemingly chaotic nature of turbulence may be underpinned by a framework of special invariant solutions of the Navier-Stokes equations.
	
	Among these, {edge states} have emerged as crucial entities in comprehending subcritical transition phenomena, prevalent in many shear flows. These are unstable solutions, whose stable manifolds form the boundary (the "edge") separating initial conditions that relaminarise from those that transition to turbulence \citep{Eckhardt2007, Avila2023}. Identifying and characterising these states, which can range from steady solutions to time-periodic or even chaotic saddles, provides a pathway to understanding the minimal seeds \citep{Pringle2012, Duguet2013} for turbulence and the geometry of the laminar-turbulent boundary \citep{Kerswell2014}.
	
	Further into the turbulent regime, the chaotic dynamics are thought to be organised around a skeleton of unstable periodic orbits (UPOs). Although trajectories in a turbulent flow are not themselves periodic, they may closely shadow various UPOs for finite times. These UPOs are exact, albeit unstable, solutions of the Navier-Stokes equations, and their properties are believed to encode essential statistical and structural features of turbulent flows \citep{Kawahara2012}. The systematic computation and analysis of UPOs thus offer a promising avenue for a more fundamental description of turbulence, moving beyond purely statistical characterisations towards a dynamical understanding. In the last years, the computational efficiency of UPOs has been significantly improved \citep{Ashtari2023}.
	
\begin{figure}
		\centering
		\includegraphics[scale=0.7]{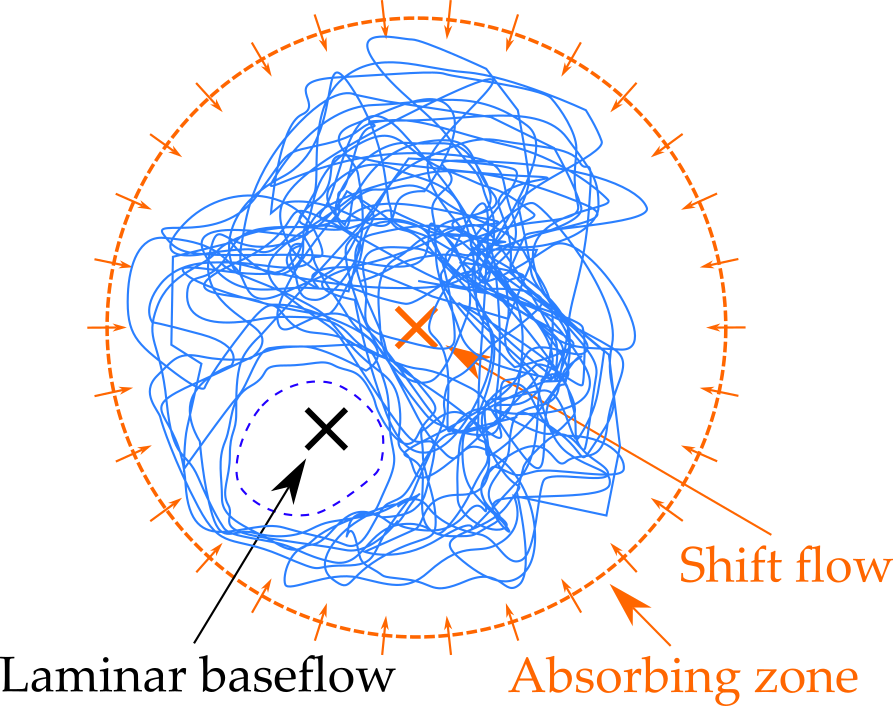} 
		\caption{The sketch of the minimal absorbing zone around the optimal shift flow}
		\label{fig:absorbing_zone}
	\end{figure}

	A rarely studied feature of shear‐flow dynamics is the \emph{absorbing zone} introduced by \citet{Dauchot1997}.  
	Using the Reynolds–Orr identity, one can show that there exists a region in the infinite‑dimensional state space such that any trajectory initiated outside this region eventually enters it and remains there.  
	Consequently, this zone contains all possible chaotic and non-chaotic attractors, periodic orbits, edge state solutions. The main objective of this study is to characterise and analyse this region in the case of fundamental flow configurations. 
	In the simplest setting the absorbing zone is a hypersphere, specified by its centre and radius.  
	The centre is termed the \emph{shift flow}, while the radius is measured by the perturbation kinetic energy.  
	A schematic illustration of this concept is provided in Figure~\ref{fig:absorbing_zone}.
	
	Assuming that the chaotic solution of a turbulent flow exhibits a random wandering behaviour in the state space, the time‑averaged (mean) flow can be approximated by the centroid of the smallest absorbing zone.  
	Below the energy‑stability threshold, denoted $\Rey_E$, the centre of the minimal absorbing zone coincides with the laminar solution and its radius collapses to zero; no attractor other than the laminar state can therefore exist.  
	However, as the Reynolds number exceeds $\Rey_E$, the size of the minimal absorbing zone increases, and additional attractors may emerge.
	Identifying minimal absorbing zones thus offers new insights into shear‑flow behaviour.  
	By employing variational equations, the shift flow can be optimised to minimise the size of the absorbing zone. 
	
Variational methods have proved highly successful across physics, and fluid mechanics is no exception.
Recently, \cite{Sanders2024} presented a Hamiltonian formulation of the Navier–Stokes equations and provided a comprehensive literature review.
Another variational approach focuses on maximising the rate of turbulent energy dissipation.
The principle was described in detail by \cite{Kerswell1999} and later solved for plane Couette flow \citep{Plasting2003} and pipe flow \citep{Plasting2004}.
Although this approach identifies the maximum dissipation—a key flow property—the resulting velocity profiles depart markedly from those observed in fully developed turbulence.
	
	In the present work we pursue an alternative route to predicting turbulent mean flows without time‑dependent simulation.  
	 Conceptually akin to a Reynolds‑averaged Navier–Stokes (RANS) model, our formulation is derived directly from the Navier–Stokes equations and requires no empirical closure.

First, the governing equations are derived in Section~\ref{sec:theory}.  
Next, a solution procedure for one‑dimensional mean flows, based on Chebyshev collocation, is described in Section~\ref{sec:solution1D}.  
The method is then validated on two canonical configurations—Couette and Poiseuille flow—and the results are reported in Section~\ref{sec:Results}.  
The minimal absorbing‑zone radius is quantified as a function of Reynolds number, and the corresponding optimal shift flows are compared with mean‑flow profiles from direct numerical simulations available in the literature.  
Finally, the main conclusions are summarised in Section~\ref{sec:Conclusion}.

\section{Theory}\label{sec:theory}

The non‑dimensional Navier–Stokes equations, written with the Einstein summation convention, are 
\begin{equation}\label{eq_NS}
	\pderb{{u}_{i}}{t}=-{u}_{j}\pderb{{u}_{i}}{x_j}-\pderb{{p}}{x_i}+\frac{1}{\Rey}\pderb{^2 {u}_i}{x_j \partial x_j} +f_i,
\end{equation}
\begin{equation}\label{eq_cont}
	\pderb{{u}_{i}}{x_i}=0.
\end{equation}
where $u_i$ and $p$ are the velocity and the pressure fields, $f_i$ is an arbitrary momentum source. 
$\Rey = L_0 U_0 / \nu$ is the Reynolds number based on  the characteristic length $L_0$, velocity  $U_0$ and kinematic viscosity $\nu$. 
	
Throughout this study the boundary conditions and momentum source are assumed to be time‑independent.
Accordingly, a stationary \emph{shift flow} $U_i$ and the associated pressure $P$ are introduced, both of which satisfy the prescribed boundary conditions and $u_i$ becomes the remaining time-dependent fluctuations around this shift flow. 
Such a decomposition resembles stability analysis, where $U_i$ usually denotes the base flow being time independent and fulfilling the governing equation. 
However, the shift flow in this analysis does not  necessarily fulfil the governing equation,  it is merely a state‑space point complying with the boundary conditions.   
After the decomposition, the following equation can be obtained
\begin{equation}\label{eq_NS_pert}
	\pderb{{u}_{i}}{t}=-{U}_{j}\pderb{{U}_{i}}{x_j}-{U}_{j}\pderb{{u}_{i}}{x_j}-{u}_{j}\pderb{{U}_{i}}{x_j}-{u}_{j}\pderb{{u}_{i}}{x_j}-\pderb{{p}}{x_i}+\frac{1}{\Rey}\pderb{^2 {U}_i}{x_j \partial x_j} + \frac{1}{\Rey}\pderb{^2 {u}_i}{x_j \partial x_j} +f_i.
\end{equation}
subject to the incompressibility constraint $\pderb{u_i}{x_i}=0$.

The non‑dimensional kinetic energy of the fluctuations is defined as
\begin{equation}\label{eq_def_e}
	e = \frac{1}{2}\int_\mathcal{V} {u}_{i}u_{i} \; \mathrm{d}\mathcal{V}.
\end{equation}
where $\mathcal{V}$ denotes the flow domain. 
The derivative of kinetic energy with respect to time is:
\begin{equation}\label{eq_de_dt}
	\tder{e}{t}= \int_\mathcal{V} \left( -u_i {U}_{j}\pderb{{U}_{i}}{x_j}-u_i{u}_{j}\pderb{{U}_{i}}{x_j}+u_i\frac{1}{\Rey}\pderb{^2 {U}_i}{x_j \partial x_j} + u_i\frac{1}{\Rey}\pderb{^2 {u}_i}{x_j \partial x_j} + u_i f_i \right)\; \mathrm{d}\mathcal{V},
\end{equation}
here certain terms have been eliminated by applying the Gauss divergence theorem and assuming periodic, far-field vanishing, or wall-bounded velocity fluctuations. These steps are analogous to those used in the proof of the well-known Reynolds-Orr identity.
A significant consequence of this simplification is that the non-linear terms ($u_j \partial u_i /\partial x_j$) do not affect the time derivative of the kinetic energy. This observation is also a contributing factor to the Sommerfeld paradox. 
Furthermore, this simplification substantially facilitates the proof of the existence of an absorbing zone.

Such a zone attracts all of the trajectories, meaning that the flow state can be outside of the zone temporarily, as it evolves in time it will eventually enter the zone and be unable to escape. Consequently, the zone contains all possible attractors. 
To prove the existence of such an absorbing zone, it is necessary to demonstrate that outside this zone, the kinetic energy decreases for any given state. In this context, the kinetic energy is similar to a Lyapunov functional.

\subsection{Infinite-amplitude perturbations}

The maximum possible growth of kinetic energy associated with perturbations of arbitrarily large amplitude about the shift flow is examined first.	
In this limit, it is convenient to investigate the temporal growth rate of the kinetic energy defined as
\begin{equation}\label{eq_def_mu_e}
	\mu_e = \frac{1}{e}\tder{e}{t}.
\end{equation}
It can be shown that, as the perturbation energy tends to infinity, the terms in (\ref{eq_de_dt}) that scale linearly with $u_i$ vanish, yielding
\begin{equation}\label{eq_mu_e_infty}
	\mu_e \to \mu_{e,\infty}= \frac{1}{e}\int_\mathcal{V}\left( -u_i{u}_{j}\pderb{{U}_{i}}{x_j}+ u_i\frac{1}{\Rey}\pderb{^2 {u}_i}{x_j \partial x_j} \right) \; \mathrm{d}\mathcal{V} \;\;\;\; \mathrm{as} \;\;\;\; e \to\infty .
\end{equation}
For such states the growth rate is independent of the perturbation amplitude; consequently, the maximum achievable growth depends solely on the spatial structure of the velocity field. The maximum of $\mu_{e,\infty}$ is obtained by solving the associated Euler--Lagrange equations,
\begin{equation}\label{eq_mu_e_infty_max}
	-u_j\left(\pderb{{U}_{i}}{x_j}+\pderb{{U}_{j}}{x_i}\right)+\frac{2}{\Rey}\pderb{{}^2{u}_{i}}{x_j\partial x_j}-\pderb{q}{x_i} = \mu_{e,\infty} u_i,
\end{equation}
where the space‑dependent field $q$ acts as a Lagrange multiplier enforcing the divergence‑free constraint on $u_i$.

Equations (\ref{eq_mu_e_infty_max}) and (\ref{eq_cont}) form an eigenvalue problem in which $\mu_{e,\infty}$ is the eigenvalue. 
The largest eigenvalue represents the maximum possible growth rate, and the corresponding eigenfunction $\tilde{u}_i$ is referred to as the critical infinite perturbation. The existence of an absorbing zone requires this eigenvalue to be negative, indicating that trajectories initialised sufficiently far from the base flow are attracted towards $U_i$. As a trajectory approaches $U_i$, its energy decreases; therefore, the linear terms neglected for large perturbations become significant, and the maximum possible growth rate increases. 

\subsection{Finite-amplitude perturbations}		
The critical energy level at which the maximal possible growth rate of the perturbation energy first becomes zero when the energy is decreased from infinity is now determined. This threshold defines the boundary of the absorbing zone. Outside this zone every trajectory must converge toward it because the growth rate is negative there. The radius of the absorbing zone is therefore obtained by evaluating the maximum of $\dd e / \dd t$ at successive energy levels while the energy is reduced and by identifying the critical level at which this maximum changes sign. The constrained maximum is determined by another Euler--Lagrange equation. The admissible velocity fields are required to be divergence‑free and to possess a prescribed energy level $\beta$:
	\begin{equation}\label{eq_def_beta_energy_level}
		 \beta - \frac{1}{2}\int_\mathcal{V} {u}_{i}u_{i} \; \mathrm{d}\mathcal{V}  = 0.
	\end{equation}
	This constraint is enforced through the Lagrange multiplier $\mu_L$, while the divergence‑free condition is imposed by the field $q$ as in the previous subsection. 
	
	The resulting Euler--Lagrange equations read
		\begin{equation}\label{eq_mu_e_max}
		-u_j\left(\pderb{{U}_{i}}{x_j}+\pderb{{U}_{j}}{x_i}\right)+\frac{2}{\Rey}\pderb{{}^2{u}_{i}}{x_j\partial x_j}-\pderb{q}{x_i} -  \mu_L u_i =U_j \pderb{U_i}{x_j} - \frac{1}{\Rey}\pderb{{}^2{U}_{i}}{x_j\partial x_j} - f_i,
	\end{equation}
	
Equations (\ref{eq_cont}), (\ref{eq_def_beta_energy_level}), and (\ref{eq_mu_e_max}) must be solved simultaneously to obtain candidate critical perturbations. The true critical state is the one whose energy \(\beta\) is maximal and for which the time derivative of the kinetic energy vanishes. Multiple solutions may exist for a given energy level because equation (\ref{eq_def_beta_energy_level}) is non-linear.

For practical implementation it is convenient to treat the Lagrange multiplier $\mu_L$ as a free parameter and to solve only the linear equations (\ref{eq_cont}) and (\ref{eq_mu_e_max}). The energy level $\beta$ then emerges as part of the solution instead of being prescribed a priori. Among the resulting states, the one satisfying $\mathrm{d}e/\mathrm{d}t = 0$ and possessing the largest $\beta$ is selected.

\begin{figure}
		\centering
		\subfigure[]{
			\includegraphics[scale=0.7]{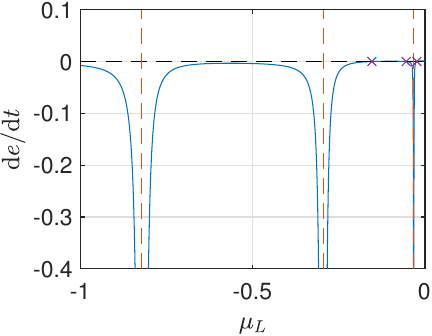} 
			\label{fig:mu_vs_edot}} 
		\hspace{5mm}
		\subfigure[]{
			\includegraphics[scale=0.7]{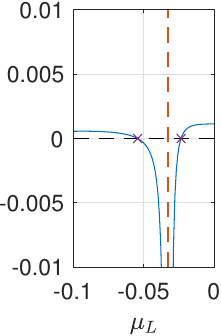}
			\label{fig:mu_vs_edot_zoom}}
		\subfigure[]{
			\includegraphics[scale=0.7]{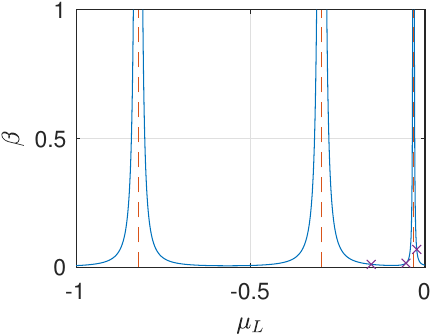} 
			\label{fig:mu_vs_beta}} 
		\hspace{5mm}
		\subfigure[]{
			\includegraphics[scale=0.7]{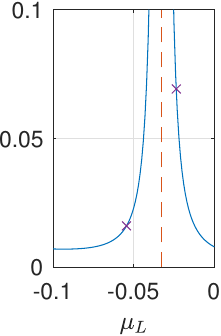}
			\label{fig:mu_vs_beta_zoom}}  
		\caption{Time derivative of the kinetic energy (a,b) and energy level (c,d) of the solutions of (\ref{eq_cont}), (\ref{eq_def_beta_energy_level}), and (\ref{eq_mu_e_max}) as functions of the Lagrange multiplier \(\mu_L\). Solutions for which \(\mathrm{d}e/\mathrm{d}t = 0\) are marked by purple crosses. Panels (b) and (d) provide zoomed views of panels (a) and (c), respectively, focusing on the neighbourhood of the critical value \(\mu_L = 0.024\). The results correspond to Poiseuille flow at \(\Rey = 200\) with the optimal shift flow.}
		\label{fig:mu_vs_edot_and_beta}
	\end{figure}

Figure~\ref{fig:mu_vs_edot_and_beta} illustrates this procedure by plotting the energy level of the solutions and the corresponding time derivative of the kinetic energy as functions of $\mu_L$. In the example shown, three values of $\mu_L$ yield $\mathrm{d}e/\mathrm{d}t = 0$. The largest energy level belongs to the largest $\mu_L$ among these values; this $\beta$ determines the size of the absorbing zone $e_\mathrm{AZ}$ around the base flow, and the associated velocity field is termed the critical finite perturbation. The other two solutions correspond to local extrema only.

It is worth mentioning that the eigenvalues \(\mu_{e,\infty}\) obtained from (\ref{eq_mu_e_infty_max}) and (\ref{eq_cont}) are also eigenvalues of the operator that appears on the left‑hand side of (\ref{eq_mu_e_max}) together with (\ref{eq_cont}). If \(\mu_L\) is set equal to one of these eigenvalues, the finite‑energy problem (\ref{eq_mu_e_max}) has no solution, whereas the energy level of the solution diverges as \(\mu_L\) approaches an eigenvalue. In figure~\ref{fig:mu_vs_edot_and_beta} the eigenvalues are indicated by vertical dashed red lines, and the associated singularities are clearly visible.

The procedure for computing the absorbing zone around a given base flow \(U_i\) is summarised below.  
\begin{enumerate}
	\item Solve (\ref{eq_mu_e_infty_max}) and (\ref{eq_cont}) to obtain the largest eigenvalue \(\mu_{e,\infty}\).  
	If \(\mu_{e,\infty} > 0\), the absorbing zone does not exist.  
	\item If \(\mu_{e,\infty} < 0\), solve (\ref{eq_mu_e_max}) and (\ref{eq_cont}) for various values of \(\mu_L\) and adjust \(\mu_L\) until $\dd e / \dd t = 0$ meaning that
	\begin{equation}\label{eq_de_dt_eq_zero}
		\int_\mathcal{V}\!\left(
		-u_i U_j\pderb{U_i}{x_j}
		-u_i u_j\pderb{U_i}{x_j}
		+\frac{1}{\Rey}\,u_i\pderb{^2 U_i}{x_j\partial x_j}
		+\frac{1}{\Rey}\,u_i\pderb{^2 u_i}{x_j\partial x_j}
		+u_i f_i
		\right)\,\mathrm{d}\mathcal{V} \;=\; 0 .
	\end{equation}
	\item Among the (possibly several) solutions that satisfy (\ref{eq_de_dt_eq_zero}), select the one with the highest energy; this energy is denoted by \(e_\mathrm{AZ}\).
\end{enumerate}

\subsection{Minimising the absorbing zone} 	
For a given flow configuration several admissible shift flows \(U_i\) may exist, each of which possesses its own absorbing zone.  
According to the present theory, the shift flow that generates the smallest absorbing zone must be selected, because it is expected to provide the best approximation to the turbulent mean flow.  
The selection can be formulated as the optimisation problem
	
\begin{equation}
	e_\mathrm{AZ, opt}  = \min_{U_i}  e_\mathrm{AZ} [U_i]
\end{equation}
	
The functional derivative of \(e_{\mathrm{AZ}}\) with respect to \(U_i\) cannot be written in explicit form; therefore, the corresponding Euler-Lagrange equation is not solved. Instead, the direct optimisation of $e_\mathrm{AZ}$ is proposed using gradient‑based methods. 
	
The convergence of such algorithms can be accelerated by exploiting the sensitivity of \(e_{\mathrm{AZ}}\) to variations of the shift flow,
\begin{equation}\label{eq_d_beta_d_Uk}
		\frac{\delta e_\mathrm{AZ}}{\delta U_i} = u_j \frac{\delta u_j}{\delta U_i}
	\end{equation}
where $\frac{\delta u_j}{\delta U_i}$ is the sensitivity of the critical finite perturbation with respect to the shift flow modification. It is obtained by perturbing equations (\ref{eq_mu_e_max}), (\ref{eq_cont}) and (\ref{eq_de_dt_eq_zero}) leading to the following equation system.

\begin{equation}\label{eq_du_dU}
	\begin{split}
\begin{bmatrix}
	A_{11} & -\left(\pder{U_1}{x_2} +\pder{U_2}{x_1} \right) & -\left(\pder{U_1}{x_3} +\pder{U_3}{x_1} \right)& -\pder{}{x_1}& -u_1 \\
	-\left(\pder{U_2}{x_1} +\pder{U_1}{x_2} \right)  &A_{22}  & -\left(\pder{U_2}{x_3} +\pder{U_3}{x_2} \right)& -\pder{}{x_2}& -u_2 \\
	-\left(\pder{U_3}{x_1} +\pder{U_1}{x_3} \right) & -\left(\pder{U_3}{x_2} +\pder{U_2}{x_3} \right)&A_{33} &  -\pder{}{x_3}& -u_3 \\
	\pder{}{x_1}& \pder{}{x_2}&\pder{}{x_3} & 0 & 0\\
	A_{51}& A_{52}&A_{53} & 0 & 0
\end{bmatrix}
\begin{bmatrix}
	\delta u_1 \vphantom{\left(\pder{U_1}{x_2}\right)} \\ \delta u_2  \vphantom{\left(\pder{U_1}{x_2}\right)} \\ \delta u_3  \vphantom{\left(\pder{U_1}{x_2}\right)} \\   \delta q \vphantom{\frac{1}{1}}\\ \delta \mu_L
\end{bmatrix}
= \\
= 
\begin{bmatrix}
	B_{11}                            & u_2\pder{}{x1} + \pder{U_1}{x_2} & u_3\pder{}{x_1} + \pder{U_1}{x_3} \\
	u_1\pder{}{x_2} + \pder{U_2}{x_1} & B_{22}                           & u_3\pder{}{x_2} + \pder{U_2}{x_3} \\
	u_1\pder{}{x_3} + \pder{U_3}{x_1} & u_2\pder{}{x_3} + \pder{U_3}{x_2}& B_{33}  \\
	0 							      &0                                 &0 \\
	B_{51}   					      &B_{52}                            &B_{53} 
\end{bmatrix}	
\begin{bmatrix}
	\delta U_1 \vphantom{\left(\pder{U_1}{x_2}\right)} \\ \delta U_2  \vphantom{\left(\pder{U_1}{x_2}\right)} \\ \delta U_3  \vphantom{\left(\pder{U_1}{x_2}\right)}
\end{bmatrix}.
\end{split}
\end{equation}
The operators appearing above are defined by
\begin{align}
	A_\mathrm{ii} &= -2\pder{U_i}{x_i} + \frac{2}{\Rey} \pder{}{x_j x_j}- \mu_L \\
	A_\mathrm{5i} &=  -{U}_{j}\pderb{{U}_{i}}{x_j}-{u}_{j}\left(\pderb{{U}_{i}}{x_j} +\pderb{{U}_{j}}{x_i} \right)+\frac{1}{\Rey}\pderb{^2 {U}_i}{x_j \partial x_j} + \frac{2}{\Rey}\pderb{^2 {u}_i}{x_j \partial x_j} + f_i \\
	B_\mathrm{11} &= u_j \pder{}{x_j} + u_1 \pder{}{x_1} +  \pder{U_1}{x_1} + U_j \pder{}{x_j} -\frac{1}{\Rey} \pderb{^2 }{x_j \partial x_j} \\
	B_\mathrm{22} &= u_j \pder{}{x_j} + u_2 \pder{}{x_2} +  \pder{U_2}{x_2} + U_j \pder{}{x_j} -\frac{1}{\Rey} \pderb{^2 }{x_j \partial x_j} \\
	B_\mathrm{33} &= u_j \pder{}{x_j} + u_3 \pder{}{x_3} +  \pder{U_3}{x_3} + U_j \pder{}{x_j} -\frac{1}{\Rey} \pderb{^2 }{x_j \partial x_j} \\
	B_\mathrm{5i} &=  -{u}_{j}\pderb{{U}_{j}}{x_i}-{U}_{j}\pderb{{u}_{i}}{x_j} +{u}_{j}\pderb{{u}_{i}}{x_i} -\frac{1}{\Rey}\pderb{^2 u_i}{x_j \partial x_j} .
\end{align}
Calculating ${\delta u_j}/{\delta U_i}$ requires inverting the operator on the left-hand side of equation (\ref{eq_du_dU}), a task that is computationally unfeasible in most cases. However, this difficulty can be overcome by directly evaluating equation (\ref{eq_d_beta_d_Uk}), where the computation can be rearranged as solving a linear equation. 

Furthermore, because the optimisation is constrained by $\mu_{e,\infty}<0$, the corresponding sensitivity can be used to accelerate convergence.  The variation of the leading eigenvalue is 
\begin{equation}
	\frac{\delta \mu_{e,\infty}}{\delta U_i} = 2 \tilde{u}_j \pder{\tilde{u}_i}{x_j} 
\end{equation} 
where $\tilde{u}_i$ is the eigenvector associated with the largest eigenvalue, namely the critical infinite perturbation. The fact that the eigenvectors are real‑valued—owing to the symmetry of the linear operator in the eigenvalue problem~\eqref{eq_mu_e_infty_max} and \eqref{eq_cont}—was invoked in the derivation of this expression. 

Using these gradients, the shift flow can be optimised to minimise the absorbing zone.

\section{Solution method for one-dimensional shift flow}\label{sec:solution1D}

The optimisation procedure is first applied to plane Couette and plane Poiseuille flows, where the base and optimal shift flows reduce to one‑dimensional wall‑normal profiles.  
Throughout, \(x_1\), \(x_2\), and \(x_3\) denote the streamwise, wall‑normal, and spanwise coordinates, respectively; the shift flows therefore depend solely on \(x_2\).

In this case, the eigenvalue problem, (\ref{eq_cont}) and (\ref{eq_mu_e_infty_max}) can be simplified by assuming a complex wave form solution
\begin{align}
	u_i(x_1, x_2, x_3) = \hat{u}_i (x_2) \mathrm{e}^{ \mathrm{i}\left(k_1 x_1 + k_3 x_3\right) }\\
	p (x_1, x_2, x_3) = \hat{p} (x_2) \mathrm{e}^{\mathrm{i}\left(k_1 x_1 + k_3 x_3\right)} .
\end{align}
with streamwise and spanwise wavenumbers \(k_1\) and \(k_3\).

The substitution yields
\begingroup
\renewcommand*{\arraystretch}{1.5}
\begin{equation}\label{eq_mu_infty_1D_matrix_form}
	\left[
	\begin{array}{cccc}
		\frac{2}{\Rey} \, L  & -\tder{U_1}{x_2}  &  0  &  -\mathrm{i} k_1   \\
		-\tder{U_1}{x_2}  & \frac{2}{\Rey}\,L  &  0  &  -D  \\
		0  &  0  & \frac{2}{\Rey}\, L  &  -\mathrm{i} k_3  \\
		\mathrm{i} k_1   &  D
		& \mathrm{i}k_3  &  0  \\
	\end{array}  \right]
	\left[
	\begin{array}{c}
		\hat{u}_1\\\hat{u}_2\\\hat{u}_3\\{\hat{q}_m}
	\end{array}  \right]
	= \tilde{\mu}_{e,\infty} 
	\left[
	\begin{array}{cccc}
		I            & 0  &  0  &  0   \\
		0			 &  I             &  0  &  0  \\
		0            &  0             &  I  &  0  \\
		0   	     &  0             &  0  &  0  \\
	\end{array}  \right]
	\left[
	\begin{array}{c}
		\hat{u}_1\\\hat{u}_2\\\hat{u}_3\\{\hat{q}_m}
	\end{array}  \right].
\end{equation}
\endgroup
where $L=D^2-(k_1^2+k_3^2)$ is the Laplace operator and $D = \tder{\square }{x_2}$ is the differential operator with respect to $x_2$, $I$ is the identity operator.

The true critical eigenvalue, ${\mu}_{e,\infty}$  is the largest one among all possible  wavenumber pairs:
\begin{equation}
	{\mu}_{e,\infty} = \max_{k_1, k_3} \tilde{\mu}_{e,\infty}.
\end{equation}
The maximum is consistently found to occur at $k_1=0$. This finding is analogous to the behaviour of the laminar base flow profile in Couette flow, as proven by \citet{Joseph1969}. 

The equation (\ref{eq_mu_e_max}), the condition for maximal growth in the case of finite perturbations, simplifies similarly. It is also a linear equation; therefore, the complex Fourier transformation can be applied. Because the right‑hand side contains only non‑oscillatory terms \((k_1=k_3=0)\), the left‑hand side forces all oscillatory \(\hat{u}_i\) to vanish, meaning that the critical finite perturbation field depends only on $x_2$. The continuity equation then implies \(u_2=u_3=0\), so that
\begin{equation}
	\frac{2}{\Rey} \pderb{^2 {u}_1}{x_2 \partial x_2} + \mu_L u_1 =  - \frac{1}{\Rey}\pderb{{}^2{U}_{1}}{x_2\partial x_2} - f_1
\end{equation}

When the shift flow coincides with the laminar base flow, the right‑hand side of the governing equation vanishes; consequently, the left‑hand side must also be zero.  All requirements for an absorbing zone are therefore satisfied except the inequality \(\mu_{e,\infty}<0\).  This condition holds whenever the Reynolds number lies below the global‑stability limit obtained from the classical Reynolds–Orr energy method, \(\Rey<\Rey_E\).  In that regime the minimal absorbing zone collapses to a single point—the laminar base‑flow state—which is consistent with the base flow being a global attractor. 
 
The equation (\ref{eq_du_dU}), which expresses the sensitivity, also reduces in one dimension to
 \begingroup
 \renewcommand*{\arraystretch}{1.5}
 \begin{equation}\label{eq_du_dU_1D}
 		\begin{bmatrix}
 			\frac{2}{\Rey}\,D^2 + \mu_L & -u_1 \\
 			\frac{1}{\Rey} \left(D^2 U_1\, +\, 2 \, D^2 u_1 \right)  + f_1   & 0  
 		\end{bmatrix}
 		\begin{bmatrix}
 			\delta u_1  \\ \delta \mu_L
 		\end{bmatrix}
 		=
 		\begin{bmatrix}
 			-\frac{1}{\Rey}D^2 \\
 			-\frac{1}{\Rey} D^2 u_1   
 		\end{bmatrix}
 		\begin{bmatrix}
 			\delta U_1  
 		\end{bmatrix}.
 \end{equation}
 \endgroup
 
Two configurations are examined: (i) plane Couette flow between moving walls, and (ii) plane Poiseuille flow driven by a constant pressure gradient between stationary walls.

The Reynolds number is defined as:
\begin{equation}
	\Rey = \frac{U_0 h }{\nu}
\end{equation}
where \(h\) is the half‑gap.  
For Couette flow \(U_0\) is half the wall‑speed difference; for Poiseuille flow it is the centreline velocity of the laminar base flow .  
The forcing term \(f_1\) vanishes for Couette flow and is chosen to yield the prescribed \(\Rey\) for Poiseuille flow.

In the numerical study \(\Rey\) was varied from \(30\) to \(500\) in increments of \(2.5\) for plane Couette flow, and from \(100\) to \(4000\) in increments of \(10\) for plane Poiseuille flow.  
At each step the optimiser is initialised with the previously converged solution. In the first step, the laminar base flow was used, however in the case of Couette flow a minor perturbation was necessary.
Optimisation was carried out in \textsc{MATLAB}\,2024b using the built‑in \texttt{fmincon} routine with the interior‑point algorithm. 
Spectral‑collocation grids containing \(71\), \(81\), and \(91\) points were tested and produced indistinguishable results; the data presented below correspond to the \(91\)-point grid.

 

\section{Results}
\label{sec:Results}
\begin{figure}
	\centering
	\subfigure[]{
		\includegraphics[scale=0.7]{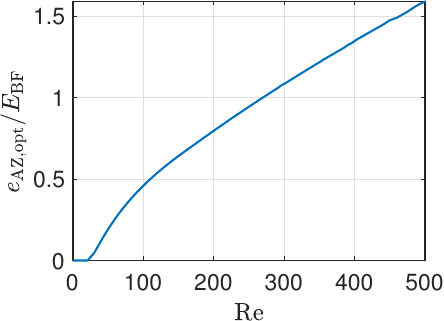}  
		\label{fig:Cou_beta_vs_Re}} 
	\hspace{5mm}
	\subfigure[]{
		\includegraphics[scale=0.7]{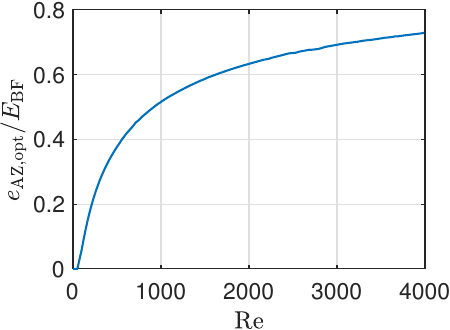}
		\label{fig:Poi_beta_vs_Re}}
	\caption{Kinetic energy of the critical finite perturbation, normalised by the kinetic energy of the laminar base flow, for (a) plane Couette and (b) plane Poiseuille flow.}
	\label{fig:beta_vs_Re}
\end{figure}

The dependence of the minimal absorbing zone on the Reynolds number is examined first.
Figure~\ref{fig:beta_vs_Re} displays the kinetic energy of the critical finite perturbation normalised by the kinetic energy of the laminar base flow.  
Below the classical energy‑stability limits, \(\Rey_E=20.6625\) for plane‑Couette flow and \(\Rey_E=49.6035\) for plane‑Poiseuille flow, the size of the region is zero. 
Immediately above these thresholds the zone expands rapidly, after which its growth reduces to an approximately linear trend in both configurations. (Unfortunately, for larger Reynolds numbers very sharp velocity profiles are obtained as optimal shift flow, which makes it difficult to analyse the further trend.)

Even at modest Reynolds numbers the energy of the critical finite perturbation remains comparable to that of the base flow.  
For Poiseuille flow the ratio stays below unity and grows only slowly, whereas in Couette flow it exceeds unity for \(\Rey\gtrsim270\) and increases more steeply.

\begin{figure}
	\centering
	\subfigure[]{
		\includegraphics[scale=0.5]{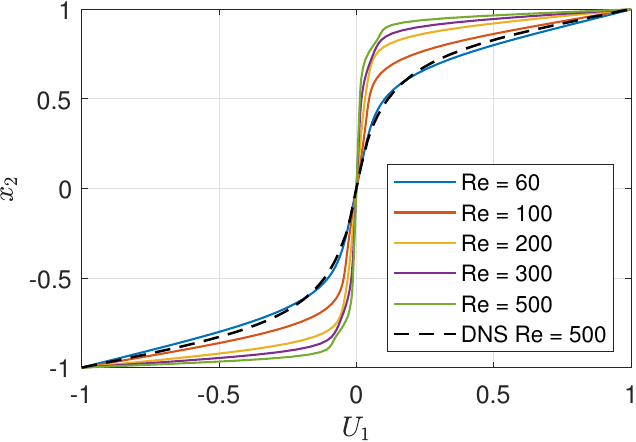} 
		\label{fig:Cou_U_shift}} 
	\hspace{5mm}
	\subfigure[]{
		\includegraphics[scale=0.5]{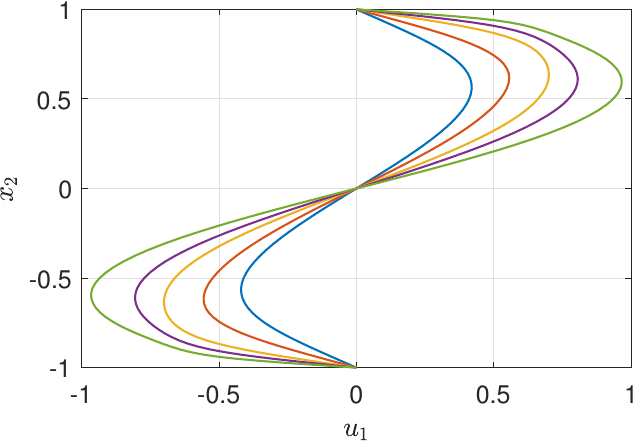}
		\label{fig:Cou_u_crit}}
	\subfigure[]{
		\includegraphics[scale=0.5]{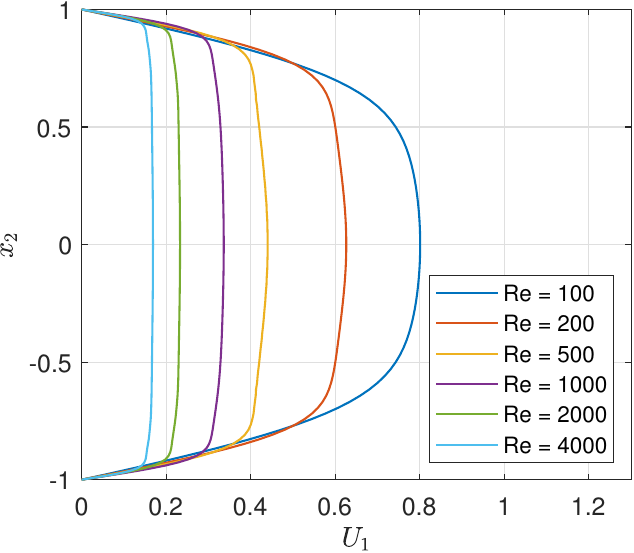} 
		\label{fig:Poi_U_shift}} 
	\hspace{5mm}
	\subfigure[]{
		\includegraphics[scale=0.5]{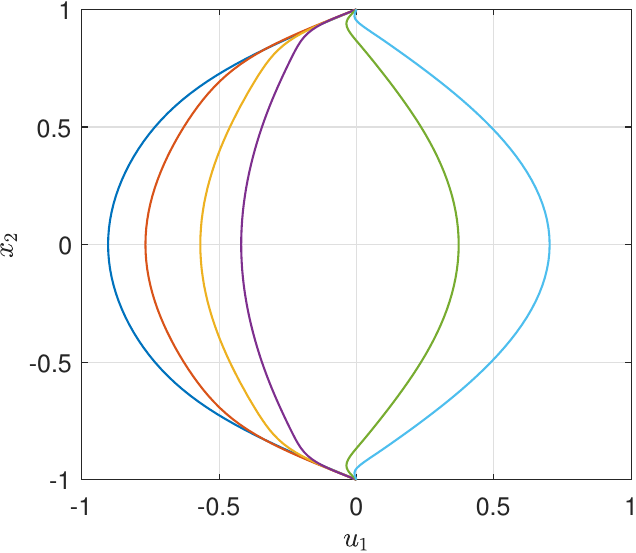}
		\label{fig:Poi_u_crit}}
	\caption{Optimal shift‐flow profiles (a,c) and their corresponding critical finite perturbations (b,d).  Panels (a–b): Couette flow; panels (c–d): Poiseuille flow.}
	\label{fig:vel_profiles_Cou_Poi}
\end{figure}

The optimal shift‑flow profiles and their corresponding critical finite perturbations are depicted in Fig.~\ref{fig:vel_profiles_Cou_Poi}. 
For Couette flow, the velocity gradient near the wall increases sharply while the slope near the channel centre approaches zero. 
This behaviour reflects the requirement that, at high perturbation amplitudes, the kinetic energy must decrease therefore viscous dissipation must dominate production. Since production is proportional to the product of fluctuation velocity and shift‑flow shear, the shear is concentrated where the fluctuations must vanish due to boundary condition: close to the walls.

In Poiseuille flow the wall gradient grows only moderately, whereas the bulk velocity is markedly reduced.  This is a consequence of the chosen non‑dimensionalisation; a Reynolds number based on bulk velocity would reveal a more Couette‑like trend.

The critical finite perturbation in Couette flow changes little in shape as \(\Rey\) increases—only its amplitude grows.  
In contrast, the Poiseuille counterpart undergoes a pronounced shape change, presumably owing to the imposed pressure gradient.

\begin{figure}
	\centering
	\subfigure[]{
		\includegraphics[scale=0.5]{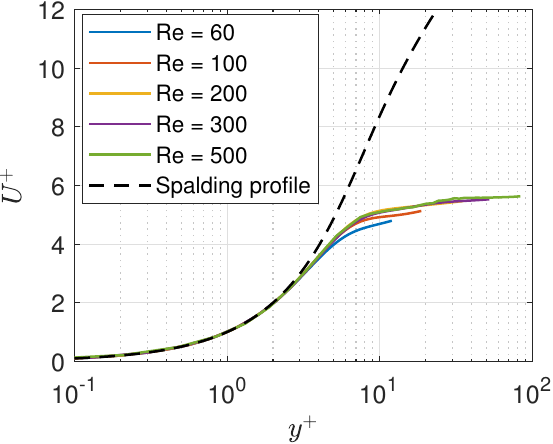}
		\label{fig:vel_profiles_turb_Cou}}
	\subfigure[]{
		\includegraphics[scale=0.5]{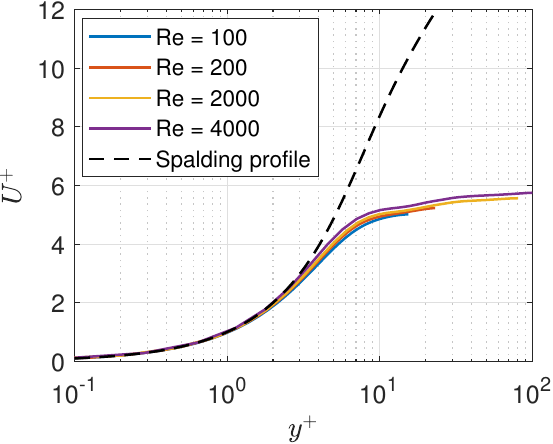}
		\label{fig:vel_profiles_turb_Poi}}		
	\caption{Velocity profiles non-dimensionalised by the friction velocity versus wall distance for (a) Couette and (b) Poiseuille flow, compared with the Spalding profile ($\kappa = 0.41, C = 5.2$).}
	\label{fig:vel_profiles_turb}
\end{figure}

The results are also compared with recent DNS simulations from the literature \citep{Cavalieri2022} at $\Rey = 500$ for Couette flow, as shown in figure \ref{fig:Cou_U_shift}. The turbulent mean profile is significantly less sharp close to the wall and does not satisfy the condition $\mu_{e,\infty} < 0$. It closely resembles the optimal shift flow profile at a much lower Reynolds number, $\Rey = 60$, although it still does not match it perfectly.

Figure~\ref{fig:vel_profiles_turb} compares the dimensionless optimal shift‑flow profiles with the classical Spalding fit for turbulent boundary layers, using viscous scaling.  
Above \(\Rey\approx200\) the optimal profiles become nearly Reynolds‑number‑independent and almost identical between the two configurations, yet they differ substantially from the turbulent law of the wall.  
Therefore, we can conclude that the approach to obtain the minimal absorbing zone cannot predict the turbulent mean profile well.

Interestingly, the optimal shift flow profiles resembles the optimal dissipation‑maximising profile obtained by \citet{Plasting2003}  (figure 6 in the cited paper) at \(\Rey=73\,000\) for the Couette flow, although the latter exhibits a slightly lower friction velocity.  There $U^+\approx4$ at $y^+ \approx 80$, while in our case $U^+\approx5.6$ at the same location.
The source of this coincidence between two quite different optimisation criteria remains unclear.

Two main factors are believed to be the main source of error in approximating the turbulent mean profile.  
First, probably the usage of kinetic energy to prove the existence of the absorbing zone simplifies the problem significantly but is not optimal. A more general energy functional—albeit harder to employ owing to the loss of the Reynolds–Orr identity—might yield a smaller zone that coincides more closely with the turbulent part of the state space.  Hopefully, such an approach could lead to the turbulent mean flow without actually averaging a time-dependent DNS solution.
Second, the assumption that the turbulent attractor explores state space randomly and uniformly is likely oversimplified; in reality the dynamics may linger near certain invariant solutions (e.g.\ unstable periodic orbits), so that the centroid of the minimal absorbing zone does not represent the mean flow.

\begin{figure}
	\centering
	\includegraphics[scale=0.5]{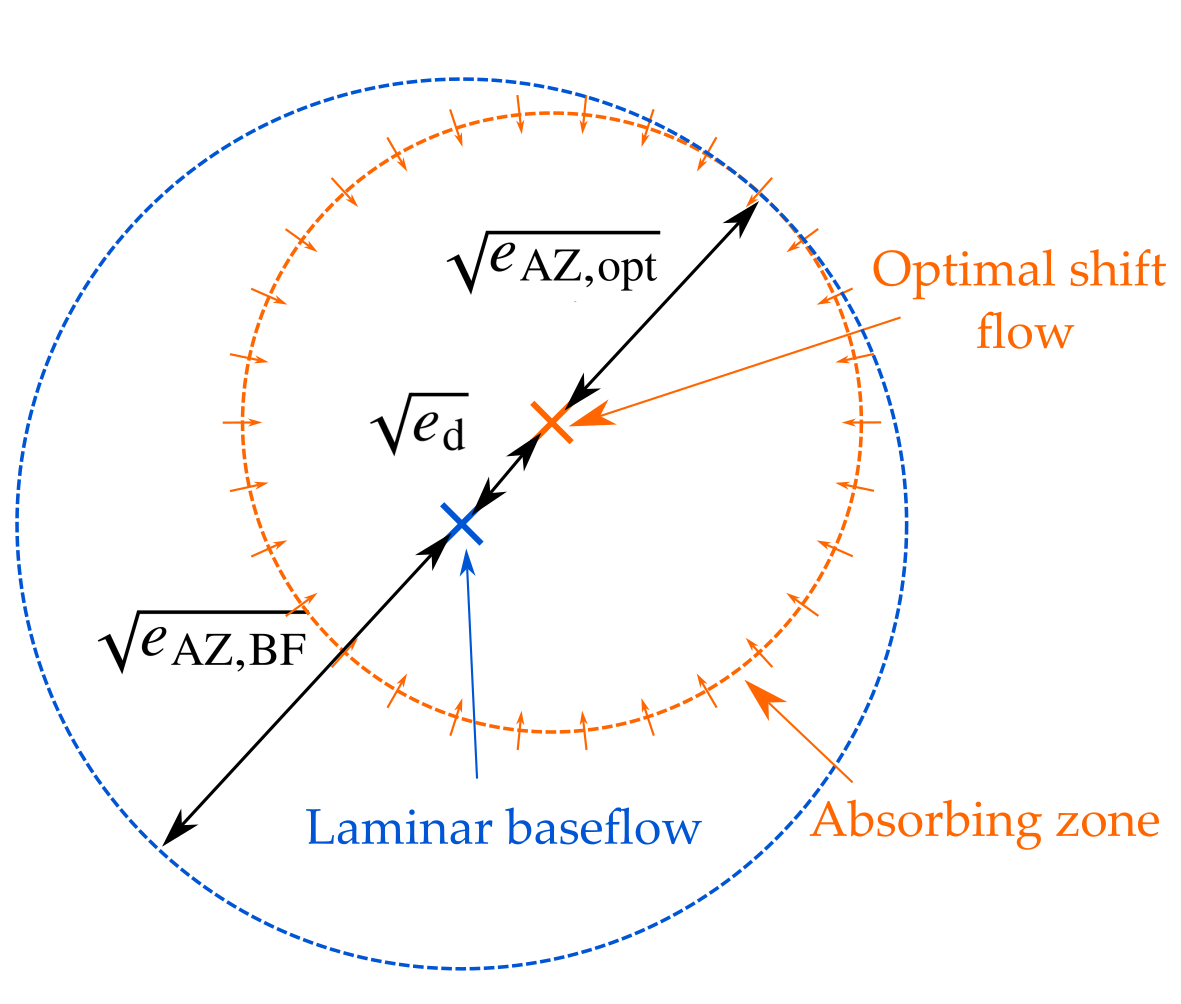} 
	\caption{The sketch of the absorbing zones around the optimal shift flow and the laminar base flow}
	\label{fig:absorbing_zone_BF}
\end{figure}

\begin{figure}
	\centering
	\subfigure[]{
		\includegraphics[scale=0.7]{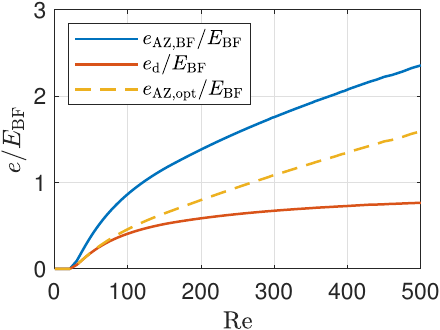} 
		\label{fig:Cou_energies_vs_Re}} 
	\hspace{5mm}
	\subfigure[]{
		\includegraphics[scale=0.7]{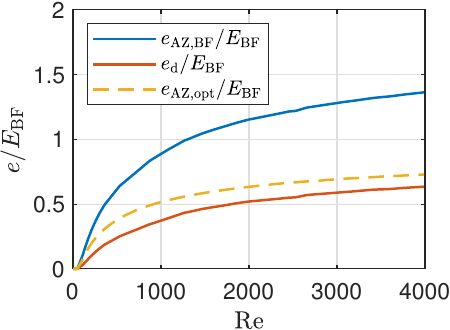}
		\label{fig:Poi_energies_vs_Re}}
	\caption{The kinetic energy of critical finite perturbations of the optimal shift flow ($e_\mathrm{AZ, opt}$), the size of the absorbing zone around the laminar base flow ($e_\mathrm{AZ, BF}$), and the energy of the difference between the base flow and the optimal shift flow ($e_\mathrm{d}$) normalised by the energy of the laminar base flow ($E_\mathrm{BF}$). (a) Couette flow, (b) Poiseuille flow.  }
	\label{fig:_energies_vs_Re}
\end{figure}

Although minimising the absorbing zone does not predict the turbulent mean profile, it offers a route to proving global stability of the laminar base flow.  
Any larger domain which includes the whole minimal absorbing zone is also an absorbing zone.
Therefore, if the laminar base flow solution is inside the minimal absorbing zone, we can also define an absorbing zone around the laminar base flow, as visualised in figure \ref{fig:absorbing_zone_BF}.
The size of this zone can be characterised by the kinetic energy:
\begin{equation}
	e_\mathrm{AZ, BF} =e_\mathrm{AZ, opt} + e_\mathrm{d}.
\end{equation}
where 
\begin{equation}
	e_\mathrm{d} = \frac{1}{2}\int_\mathcal{V} \left({U}^\mathrm{BF}_{i} - {U}^\mathrm{SF}_{i} \right)\left({U}^\mathrm{BF}_{i} - {U}^\mathrm{SF}_{i} \right) \; \mathrm{d}\mathcal{V},
\end{equation}
is the kinetic energy of the difference between the laminar base flow \(U_i^{\mathrm{BF}}\) and the optimal shift flow \(U_i^{\mathrm{SF}}\). 
These energies, normalised by the laminar kinetic energy \(E_{\mathrm{BF}}\), are plotted in figure~\ref{fig:_energies_vs_Re}.  
First, it can be seen that $e_\mathrm{d}\le e_\mathrm{AZ, BF} $ in both flow configurations; therefore, the minimal absorbing zone includes the laminar state, as expected, since in both configurations, the base flow is linearly stable.
Consequently, there exists an absorbing zone around the laminar base state.

This absorbing zone can be utilised to prove the global stability of the base flow.
In the case of a linearly stable base state, there exists a region of attraction of the laminar state.
If the provable region of attraction is large enough and this minimal zone is fully inside, it proves the global stability of the flow.
Recently, for finite-dimensional systems, \cite{Nagy2025} showed an efficient way to construct such a conditional Lyapunov function to calculate the size of the region of attraction and showed possible ways of extension to infinite-dimensional systems.
However, it must be mentioned that even in these preliminary results, the normalised allowable kinetic energy is of the order of $10^{-2}-10^{-3}$.
Since the size of the minimal absorbing zone increases rapidly after the well-known stability limit $\Rey_E$ and is of the order of $1$, further improvement is still necessary to significantly increase the provable global stability limit of shear flows.
 
\begin{figure}
	\centering
	\includegraphics[scale=0.7]{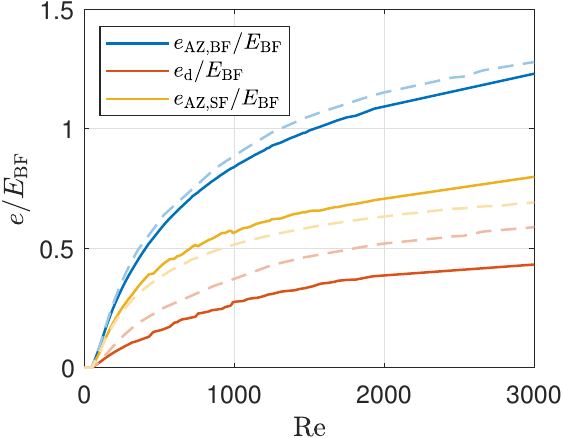}
	\caption{Poiseuille flow: comparison of the original optimisation (dashed, pale curves) with direct minimisation of \(e_{\mathrm{AZ,BF}}\) (solid curves).  Quantities as in figure~\ref{fig:_energies_vs_Re}.}
	\label{fig:_energies_vs_Re_opt_BF}
\end{figure}

A further attempt is made to slightly improve the global stability results.
In this case, the size of the absorbing zone around the laminar base flow ($e_\mathrm{AZ, BF}$) is minimised instead of only the size of the absorbing zone.
This needs only a minor, trivial modification in the calculation of the cost function and the gradient.
In the case of Couette flow, the results were identical to the previous optimisation procedure; therefore, these results are not shown.
At the same time, a non-negligible improvement is made in the case of Poiseuille flow.
The observed difference between the flows is attributed to the presence of forcing, which is zero in the case of Couette flow.
The result of the new optimisation procedure is presented in figure \ref{fig:_energies_vs_Re_opt_BF}.
Compared with the previous optimisation, the absorbing zone around the laminar state is $\approx 5\%$ smaller. In contrast, the absorbing zone around the shift flow grew by $\approx 10\%$, and the size difference between the shift flow and the laminar base flow decreased by $\approx 35\%$.

\section{Conclusion}
\label{sec:Conclusion}

In this study, a variational method was introduced and applied to determine the minimal absorbing zone for incompressible shear flows, exploiting the kinetic‐energy balance and the Reynolds-Orr identity.
The primary hypothesis was that the centroid of this minimal zone—the optimal shift flow—could provide a direct approximation of the turbulent mean flow without requiring time-dependent simulations or empirical closures.

The procedure was successfully implemented for plane Couette and plane Poiseuille flows, producing optimised shift–flow profiles and corresponding absorbing–zone sizes over a range of Reynolds numbers. However, comparison with direct numerical simulation data and canonical turbulent profiles revealed substantial quantitative discrepancies. The optimal shift flows systematically exhibit steeper near‑wall gradients than their turbulent counterparts, indicating that the hypothesis, in its present form, is not supported.

Two plausible sources of this discrepancy are suggested. First, reliance on the standard kinetic energy, while analytically convenient (it excludes explicit contributions from the nonlinear advection terms), may be overly restrictive for delineating the true boundary of the absorbing zone. Employing a more general energy functional could alleviate this limitation, albeit at the cost of a much more challenging optimisation problem. Second, the assumption that a chaotic turbulent trajectory explores the state space randomly and uniformly is likely an oversimplification; the dynamics may instead stay near invariant solutions, such as unstable periodic orbits. Therefore, the centroid of the minimal absorbing zone need not coincide with the turbulent mean state.

Despite the failure of the primary hypothesis, the methodology retains significant value for global‐stability analysis. The computed minimal absorbing zone constitutes a rigorous, provable subset of state space that must contain all attractors, including the laminar solution. An absorbing region centred on the laminar base flow was also constructed. If this region can be shown to lie entirely within the laminar state’s region of attraction, global stability follows. Although the present bounds remain too large to markedly extend existing global‑stability limits, the approach is novel and promising. Future work based on more general Lyapunov functionals may yield smaller, tighter absorbing zones, potentially opening a new route to estimating turbulent mean states and deepening the understanding of hydrodynamic stability.

	\backsection[Funding]{This paper was supported by the János Bolyai Research Scholarship of the Hungarian Academy of Sciences.}
	
	\backsection[Declaration of interests]{The author reports no conflict of interest.}

	\backsection[Author ORCIDs]{P. T. Nagy, https://orcid.org/0000-0002-8024-3824
	}

	\appendix

\bibliographystyle{jfm}
\bibliography{biblography_V1}

\begin{thebibliography}{17}
\expandafter\ifx\csname natexlab\endcsname\relax\def\natexlab#1{#1}\fi
\def\au#1{#1} \def\ed#1{#1} \def\yr#1{#1}\def\at#1{#1}\def\jt#1{\textit{#1}}
  \def\bt#1{#1}\def\bvol#1{\textbf{#1}} \def\vol#1{#1} \def\pg#1{#1}
  \def\publ#1{#1}\def\arxiv#1{#1}\def\org#1{#1}\def\st#1{\textit{#1}}

\bibitem[Ashtari \& Schneider(2023)]{Ashtari2023}
{\sc \au{Ashtari, Omid} \& \au{Schneider, Tobias~M.}} \yr{2023}
  \at{Identifying invariant solutions of wall-bounded three-dimensional shear
  flows using robust adjoint-based variational techniques}.  \jt{Journal of
  Fluid Mechanics}  \bvol{977},  \pg{A7}.

\bibitem[Avila {\em et~al.\/}(2023)Avila, Barkley \& Hof]{Avila2023}
{\sc \au{Avila, Marc}, \au{Barkley, Dwight} \& \au{Hof, Björn}} \yr{2023}
  \at{Transition to turbulence in pipe flow}.  \jt{Annual Review of Fluid
  Mechanics}  \bvol{55}~(Volume 55, 2023),  \pg{575--602}.

\bibitem[Cavalieri \& Nogueira(2022)]{Cavalieri2022}
{\sc \au{Cavalieri, Andr\'e V.~G.} \& \au{Nogueira, Petr\^onio A.~S.}}
  \yr{2022}  \at{Reduced-order {G}alerkin models of plane {C}ouette flow}.
  \jt{Phys. Rev. Fluids}  \bvol{7},  \pg{L102601}.

\bibitem[Doering(2009)]{Doering2009}
{\sc \au{Doering, Charles~R.}} \yr{2009}  \at{The {3D} {N}avier-{S}tokes
  {P}roblem}.  \jt{Annual Review of Fluid Mechanics}  \bvol{41}~(Volume 41,
  2009),  \pg{109--128}.

\bibitem[Duguet {\em et~al.\/}(2013)Duguet, Monokrousos, Brandt \&
  Henningson]{Duguet2013}
{\sc \au{Duguet, Yohann}, \au{Monokrousos, Antonios}, \au{Brandt, Luca} \&
  \au{Henningson, Dan~S.}} \yr{2013}  \at{{Minimal transition thresholds in
  plane Couette flow}}.  \jt{Physics of Fluids}  \bvol{25}~(8).

\bibitem[Eckhardt {\em et~al.\/}(2007)Eckhardt, Schneider, Hof \&
  Westerweel]{Eckhardt2007}
{\sc \au{Eckhardt, Bruno}, \au{Schneider, Tobias~M.}, \au{Hof, Bjorn} \&
  \au{Westerweel, Jerry}} \yr{2007}  \at{Turbulence transition in pipe flow}.
  \jt{Annual Review of Fluid Mechanics}  \bvol{39}~(Volume 39, 2007),
  \pg{447--468}.

\bibitem[Jiménez(2018)]{Jimenez2018}
{\sc \au{Jiménez, Javier}} \yr{2018}  \at{Coherent structures in wall-bounded
  turbulence}.  \jt{Journal of Fluid Mechanics}  \bvol{842},  \pg{P1}.

\bibitem[Joseph \& Carmi(1969)]{Joseph1969}
{\sc \au{Joseph, D.~D.} \& \au{Carmi, S}} \yr{1969}  \at{{Stability of
  Poiseuille flow in pipes, annuli, and channels}}.  \jt{Quarterly of Applied
  Mathematics}  \bvol{26}~(4),  \pg{575--599}.

\bibitem[Kawahara {\em et~al.\/}(2012)Kawahara, Uhlmann \& van
  Veen]{Kawahara2012}
{\sc \au{Kawahara, Genta}, \au{Uhlmann, Markus} \& \au{van Veen, Lennaert}}
  \yr{2012}  \at{The significance of simple invariant solutions in turbulent
  flows}.  \jt{Annual Review of Fluid Mechanics}  \bvol{44}~(Volume 44, 2012),
  \pg{203--225}.

\bibitem[Kerswell(1999)]{Kerswell1999}
{\sc \au{Kerswell, R.~R.}} \yr{1999}  \at{Variational principle for the
  {N}avier-{S}tokes equations}.  \jt{Phys. Rev. E}  \bvol{59},
  \pg{5482--5494}.

\bibitem[Kerswell {\em et~al.\/}(2014)Kerswell, Pringle \&
  Willis]{Kerswell2014}
{\sc \au{Kerswell, R~R}, \au{Pringle, C C~T} \& \au{Willis, A~P}} \yr{2014}
  \at{An optimization approach for analysing nonlinear stability with
  transition to turbulence in fluids as an exemplar}.  \jt{Reports on Progress
  in Physics}  \bvol{77}~(8),  \pg{085901}.

\bibitem[Nagy(2025)]{Nagy2025}
{\sc \au{Nagy, Péter~Tamás}} \yr{2025}  \at{Constructing quadratic {L}yapunov
  functions for conditionally stable fluid dynamics systems}.  \jt{Journal of
  Fluid Mechanics}  \bvol{1014},  \pg{A25}.

\bibitem[{Olivier Dauchot} \& {Paul Manneville}(1997)]{Dauchot1997}
{\sc \au{{Olivier Dauchot}} \& \au{{Paul Manneville}}} \yr{1997}  \at{Local
  $versus$ global concepts in hydrodynamic stability theory}.  \jt{J. Phys. II
  France}  \bvol{7}~(2),  \pg{371--389}.

\bibitem[Plasting \& Kerswell(2003)]{Plasting2003}
{\sc \au{Plasting, S.~C.} \& \au{Kerswell, R.~R.}} \yr{2003}  \at{Improved
  upper bound on the energy dissipation rate in plane {C}ouette flow: the full
  solution to {B}usse’s problem and the {C}onstantin–{D}oering–{H}opf
  problem with one-dimensional background field}.  \jt{Journal of Fluid
  Mechanics}  \bvol{477},  \pg{363–379}.

\bibitem[Plasting \& Kerswell(2004)]{Plasting2004}
{\sc \au{Plasting, S.~C.} \& \au{Kerswell, R.~R.}} \yr{2004}  \at{A friction
  factor bound for transitional pipe flow}.  \jt{Physics of Fluids}
  \bvol{17}~(1),  \pg{011706},  \arxiv{arXiv:
  https://pubs.aip.org/aip/pof/article-pdf/doi/10.1063/1.1828103/14691402/011706\_1\_online.pdf}.

\bibitem[Pringle {\em et~al.\/}(2012)Pringle, Willis \& Kerswell]{Pringle2012}
{\sc \au{Pringle, Chris~C. T.}, \au{Willis, Ashley~P.} \& \au{Kerswell,
  Rich~R.}} \yr{2012}  \at{Minimal seeds for shear flow turbulence: using
  nonlinear transient growth to touch the edge of chaos}.  \jt{Journal of Fluid
  Mechanics}  \bvol{702},  \pg{415–443}.

\bibitem[Sanders {\em et~al.\/}(2024)Sanders, DeVoria, Washuta, Elamin, Skenes
  \& Berlinghieri]{Sanders2024}
{\sc \au{Sanders, John~W.}, \au{DeVoria, A.C.}, \au{Washuta, Nathan~J.},
  \au{Elamin, Gafar~A.}, \au{Skenes, Kevin~L.} \& \au{Berlinghieri, Joel~C.}}
  \yr{2024}  \at{A canonical {Hamiltonian} formulation of the {Navier–Stokes}
  problem}.  \jt{Journal of Fluid Mechanics}  \bvol{984},  \pg{A27}.

\end{thebibliography}

\end{document}